\documentclass[oldversion]{aa}  
\usepackage{graphicx}
\usepackage{lscape}
\usepackage{txfonts}
\usepackage{natbib} 
\bibpunct{(}{)}{;}{a}{}{,} 

\begin{document}

\title{On the kinematic age of brown dwarfs:\\ Radial velocities and space motions of 43 nearby L dwarfs}
\titlerunning{Space motions of brown dwarfs}

\author{A. Seifahrt\inst{1,2}
  \and
  A. Reiners\inst{1}
  \and
  K.~A.~M. Almaghrbi\inst{1}
  \and 
  G. Basri\inst{3}
}

\institute{
 Universit\"at G\"ottingen, Institut f\"ur Astrophysik, Friedrich-Hund-Platz 1, D-37077 G\"ottingen, Germany
  \and
 Physics Department, University of California, Davis, CA 95616, USA\\  \email{seifahrt@physics.ucdavis.edu}
\and
  Astronomy Department, University of California, Berkeley, CA 94720, USA 
  }

\date{as of \today}


\abstract
{ We present radial velocity measurements of a sample of L0--L8 dwarfs
  observed with VLT/UVES and Keck/HIRES. We combine these measurements
  with distance and proper motion from the literature to determine
  space motions for 43 of our targets. We identify nine candidate
  members of young moving groups, which have ages of 50--600~Myr according to their
  space motion.  From the total velocity dispersion of the 43 L dwarfs,
  we calculate a kinematic age of $\sim5$~Gyr for our sample. This age
  is significantly higher than the $\sim3$~Gyr age known for late M
  dwarfs in the solar neighbourhood. We find that the distributions of
  the $U$ and $V$ velocity components of our sample are clearly
  non-Gaussian, placing the age estimate inferred from the full space motion
  vector into question. The $W$-component exhibits a distribution more
  consistent with a normal distribution, and from $W$ alone we derive
  an age of $\sim$3\,Gyr, which is the same age found for late-M dwarf 
  samples. Our brightness-limited sample is probably contaminated by a 
  number of outliers that predominantly bias the $U$ and $V$ velocity
  components. The origin of the outliers remain unclear, but we
  suggest that these brown dwarfs may have gained their high velocities
  by means of ejection from multiple systems during their formation.  }

\keywords{Stars: low-mass, brown dwarfs -- Techniques: radial velocities}

\maketitle
%

\section{Introduction}

Early- and mid-L dwarfs encompass a mass range that includes very
low mass stars as well as more massive brown dwarfs around the
hydrogen-burning minimum mass
\citep[HBMM$\approx0.07$--$0.08~\mathrm{M}_\odot$,][]{Chabrier97}. The
kinematics of these stars, especially within the solar
neighbourhood, is important because the intrinsic velocity
dispersion of the stars can be used to determine their age
\citep{Wielen77,Fuchs01}. Brown dwarfs may then serve in turn as
Galactic chronometers \citep{Burgasser09}, given that they constantly
cool over time and never reach a stable state on the main sequence.

While an age of $\sim3$~Gyr for the nearby late-M dwarfs is now well
established \citep{Reid02a, RB09}, kinematic ages of L and T dwarfs
remain preliminary. \citet{Zapatero07} derive an age $\sim1$~Gyr
for 21 L and T dwarfs based on space motions of these objects, while
\citet{Faherty09} derive an age of 3--6~Gyrs for a large sample of L
and T dwarfs, based on their distance and proper motion. The
discrepancy between these results is probably due to differences
between the age computations but may also be attributed to the small
number of objects with measured radial velocities, and thus, 3D space
motions.

Kinematic studies of brown dwarfs have been predominantly based on
proper motion measurements \citep[see, e.g.,
][]{Schmidt07, Jameson08, Casewell08, Faherty09}. Absolute radial
velocities of brown dwarfs have been obtained for small samples by, e.g.,
\citet{B00}, \citet{BJ04}, \citet{Blake07}, or \citet{Zapatero07}. In this
paper, we measure radial velocities and provide space motions for 43
L0--L8 dwarfs, thus for a much larger sample than obtained before.

\section{Sample and analysis}\label{sec:sample}

\citet[][hereafter RB08]{RB08} present high resolution ($R \ga
30,000$) optical spectra of 45 L dwarfs\footnote{One of the targets
  (2MASS J1159+00) has too low a signal-to-noise ratio to compute a
  robust radial velocity, reducing the initial sample size to 44. One
  more target (2MASS J1854+84) has no published proper motion and
  distance, reducing the number of final targets in our sample to 43.} obtained with
VLT/UVES and Keck/HIRES. They measure rotational velocities and
chromospheric activity to investigate rotational braking in M and L
dwarfs. The sample selection and observations are presented in Sect.~2 
of RB08, the targets being essentially the brightest L dwarfs
distributed over the whole sky \citep{Martin99, Kirkpatrick00, Cruz03, Kendall04, Reid08}. 
No preselection of any other parameter was made.

We use Gl~406 as our primary radial velocity standard \citep[M6V,
$v_\mathrm{rad} = 19\pm1$km\,s$^{-1}$,][]{MB03} and employ a cross-correlation
technique to measure the radial velocity of each star for various
spectral orders (see, e.g., \citealt{MB03} and \citealt{B00} for a detailed description of the method). 
We focused on the resonance lines of \ion{Cs}{I} and
\ion{Rb}{I} as well as the prominent features of TiO, VO and, most
importantly, FeH. Telluric absorption lines of the O$_2$ A-band as
well as strong H$_2$O lines around $\lambda\lambda$ 930--960 nm were
used to correct for shifts in the wavelength solution of each spectrum
with respect to our radial velocity standard. We followed the
conservative approach of adopting the standard deviation of the radial
velocities obtained in different spectral orders as the intrinsic
error of our final radial velocities. The intrinsic uncertainties
range from 0.2~km\,s$^{-1}$ for the earliest L dwarfs up to 4~km\,s$^{-1}$ for the
latest and most rapidly rotating targets. The final uncertainty in
most cases is dominated by the uncertainty in the absolute radial
velocity of Gl406, which we added in quadrature to the intrinsic
uncertainty of each target. All radial velocities were finally
corrected for barycentric motion.

\begin{figure}
  \centering
  \resizebox{\hsize}{!}{\includegraphics[]{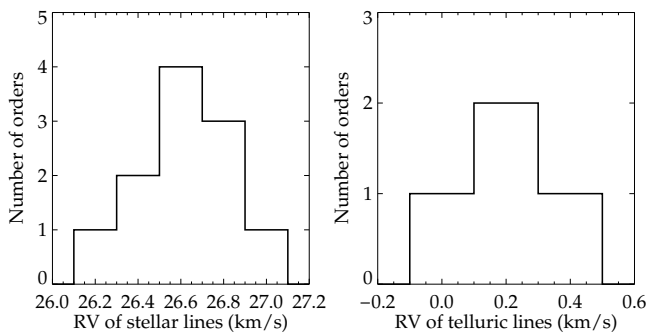}}
  \caption{\label{fig:xcorr}Example of the distribution of radial
    velocities obtained in different spectral orders from stellar
    lines (left) and telluric lines (right) in the spectrum of
    2M1645-1319 (L1.5V). See text for details.  }
\end{figure}

As an example, we show in Fig.~\ref{fig:xcorr} the results obtained
for the L1.5 dwarf 2M1645-1319.
The radial velocities obtained from stellar lines in 11 different
orders show a nearly Gaussian distribution with a mean of 26.6~km\,s$^{-1}$
and a standard deviation of 0.21~km\,s$^{-1}$. The telluric lines in 4
different orders show an offset of $0.2\pm0.16$~km\,s$^{-1}$. Hence, we
determined a radial velocity of $26.4\pm1.0$~km\,s$^{-1}$.

For all L dwarfs later than L2, we also used the spectra of
2MASS1645-1319 (L1.5) and 2MASS1045-0149 (L1.0) as secondary radial
velocity standards. Absolute radial velocities for both objects were
determined against Gl406 with a precision of $\sim0.3$~km\,s$^{-1}$. Both
targets are relatively slow rotators ($v\sin{i}=9$~km\,s$^{-1}$ and $<$3~km\,s$^{-1}$,
respectively, RB08) and their respective VLT/UVES and Keck/HIRES
spectra have a high signal-to-noise ratio, making them well suited as
cross-correlation templates. Radial velocities of all targets measured
relative to Gl406 and our L dwarf templates are always identical, but
the latter correlation produced more robust results with smaller
uncertainties since the L dwarf templates are a closer match to the 
spectral features of the targets. Thus, we adopted the radial velocities obtained from
the cross-correlation with our secondary velocity standards for all
targets later than L2 as our final results.

\section{Results}

\subsection{Radial velocities}
We present the final results from our radial velocity measurements in
Table~\ref{maintable}. Twelve targets were observed with VLT/UVES and
Keck/HIRES at different epochs. In all cases, the radial
velocities obtained from both runs were fully consistent within their
individual uncertainties, showing no sign of significant radial
velocity variability on a km\,s$^{-1}$ level. We also compared our
results to radial velocities reported in the literature
\citep{B00,BJ04,Zapatero07,Blake07}. We found good agreement with our
values in 16 cases.  However, we found two cases, namely 2M0255-47 and
2M1305-25, where our radial velocities are inconsistent with other
measurements.  We discuss these objects in Sect.~\ref{sec:notes}.

\subsection{Space motions}\label{sec:UVW} 
To calculate the space motions of our sample, we combine our radial
velocity measurements with distance and proper motion data collected 
from the literature, mainly from the extensive database of
\citet{Faherty09}. Distances are based mostly on spectrophotometric
estimates and have a typical uncertainty of 1~pc. A few L dwarfs in
our sample have measured trigonometric parallaxes, which notably
reduces their distance errors. We compared the reported
proper motions and distances in \citet{Faherty09} with other sources
and used the values with the smallest uncertainties. Details are
provided in Table~\ref{maintable}.

Space motions were computed using the IDL routine
\texttt{gal\_uvw}\footnote{Written by W.~Landsman and S.~Koposov,
  based on \citet{JS87}.}. All space motions are relative to the sun.
We adopt a sign convention for $U$ that is positive towards the
Galactic Center. We determine errors in distance, proper motion, and
radial velocity in a Monte Carlo fashion to compute error estimates in
$U$,$V$, and $W$. In this computation, we derive space motions for all
possible combinations of input parameters and their $1\sigma$
uncertainties. We then calculate the standard deviation in $U$,$V$, and
$W$ for each target from this distribution and adopt these values as
the $1\sigma$ uncertainties in $U$,$V$, and $W$.  The calculated space
motions and their respective errors are given in Table~\ref{maintable} along
with the distance and proper motion data obtained from the literature.

\subsection{Membership in young moving groups}\label{sec:mg}
Young moving groups are associations of coeval and comoving stars.
Members of moving groups can be identified either on the basis of proper motion
and position on the sky, using the moving cluster method \citep[see,
e.g.,][]{Bannister07} or by comparing the 3D space motions of individual objects
with the mean space motion of the moving group. We found members of
four young moving groups (MG), namely of the Hyades, the Pleiades MG, 
the AB Dor MG, and the UMa MG, with ages ranging from 50--600~Myr. 
All these moving groups are close-by and located in or near the velocity 
ellipsoid of the young (thin) disk \citep[$-50<U<+20$~km\,s$^{-1}$, $-30<V<0$~km\,s$^{-1}$, and
$-25<W<+10$~km\,s$^{-1}$; see, e.g.,][and references therein]{Leggett92}.
No members were found for other young and close moving groups 
considered in our analysis, such as TW Hydrae Association, the $\beta$ Pictoris MG, 
and the Tucana/Horologium Association \citep[see, e.g.,][]{Zuckerman04}. While 
their $UVW$ velocity ellipsoids are close to those of the Pleiades MG
and the AB Dor MG, we excluded memberships based on distance or
position on the sky in all cases where confusion in the $UVW$ space limited
our ability to assign memberships based on space motions only. 

\begin{figure*}
  \centering
  \resizebox{\hsize}{!}{
    \includegraphics[]{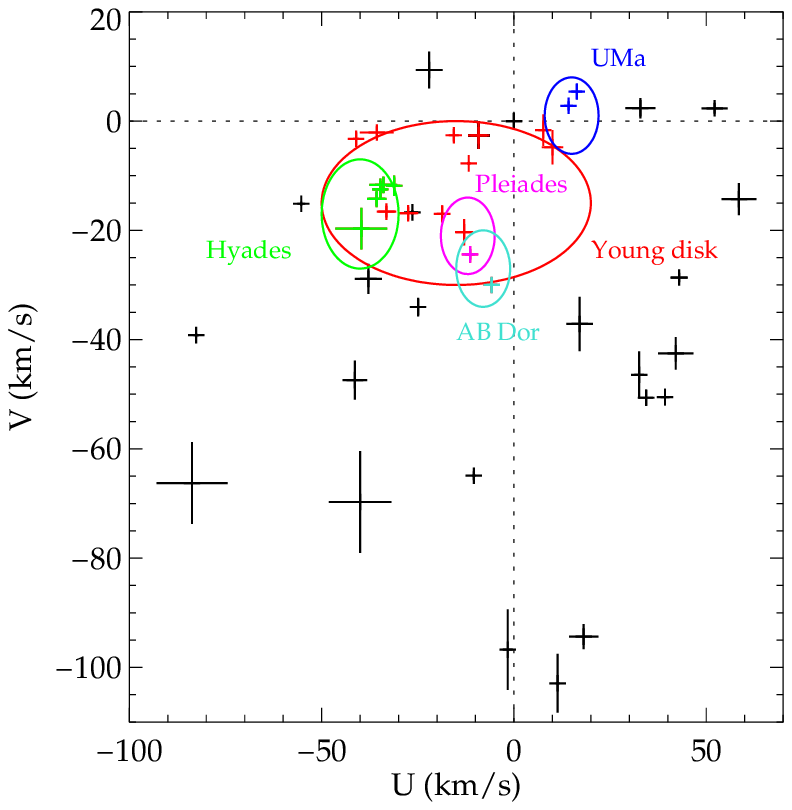}
    \includegraphics[]{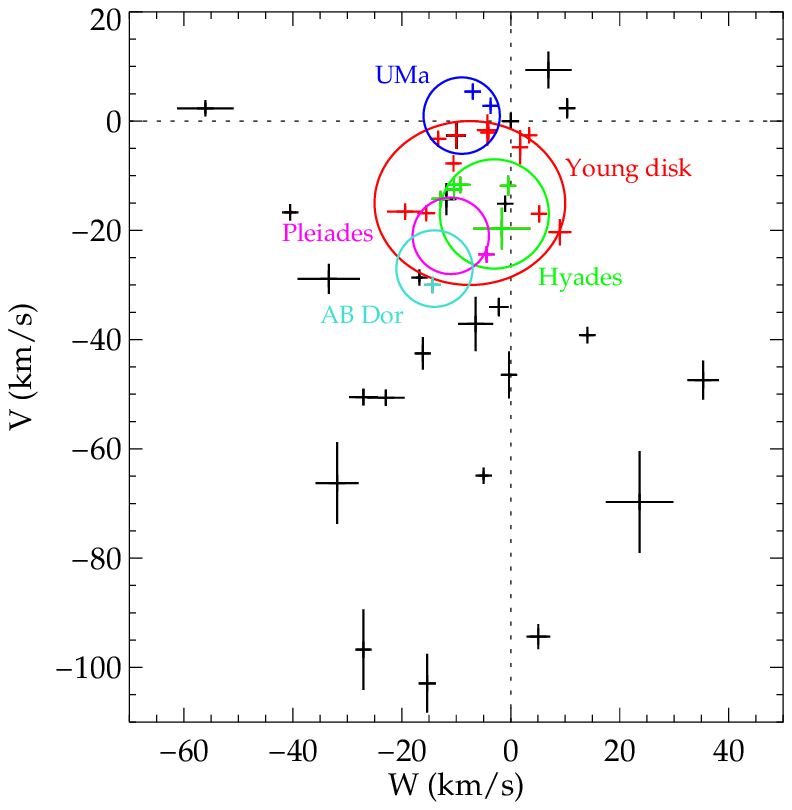}
  }
  \caption{\label{fig:UVW}Space velocities in the $U$--$V$ (left) and
    $W$--$V$ (right) plane. The velocity ellipsoids for the young disk
    as well for some young moving groups are shown. Moving groups 
    considered in the analysis but without members in our sample are 
    not shown for clarity. Objects in colour are candidate members of 
    the respective moving groups, as indicated in Table~\ref{maintable}.}
\end{figure*}

The space motion of our targets are shown in Fig.~\ref{fig:UVW}. We
overplot the velocity ellipsoids of the young moving groups in the
same figure, adopting a typical dispersion of 10~km\,s$^{-1}$ for the Hyades
and 7~km\,s$^{-1}$ for the Pleiades, AB Dor, and UMa in accordance with
\citet{Chen97} and \citet{Chereul99}. We compare the computed space
motion of our targets to the typical space motion of both the young disk
and the young moving groups listed in \citet{Zuckerman04}. We 
found a total of 11 objects that belong to the galactic young disk population (see
Table~\ref{maintable}) and another 9 objects that belong to the young
moving groups (see Table~\ref{table:mvg}). Comments on individual
targets are given in Sect.~\ref{sec:notes}.

We note that only five out of 43 targets ($\sim12$\%) are candidate
members of the Hyades. This is in notable contrast to the $\sim40$\%
(8 out of 21 targets) of L and T dwarfs reported by \citet{Zapatero07}
that are within or close to the 2~$\sigma$ ellipsoid of the Hyades. While 
this membership criterion is weaker than that applied by ourselves, the space
motion of our sample have a much wider dispersion in $UVW$ than the
sample of \citet{Zapatero07} and significantly fewer objects fall in
or near the velocity ellipsoid of the Hyades in our case.

\begin{table}
\caption{Candidate members of young moving groups based on our $UVW$ analysis}             
\label{table:mvg}      
\centering          
\begin{tabular}{l c c}
\hline\hline       
Moving group$^a$ & $UVW$ (km\,s$^{-1}$) & Age (Myr) \\
\hline
\noalign{\smallskip}
Hyades Cluster & $-40,-17,-3$ & $\sim600$ \\
\hline
\noalign{\smallskip}
\multicolumn{3}{l}{2MASS J02550357-4700509}\\
\multicolumn{3}{l}{2MASS J10452400-0149576$^b$}\\
\multicolumn{3}{l}{2MASS J10473109-1815574$^b$}\\
\multicolumn{3}{l}{2MASS J14413716-0945590}\\
\multicolumn{3}{l}{2MASS J22000201-3038327}\\

\noalign{\smallskip}
\hline\noalign{\smallskip}
Ursa Majoris MG & $+14,+1,-9$ & $\sim300$ \\
\hline
\noalign{\smallskip}
\multicolumn{3}{l}{2MASS J03140344+1603056}\\
\multicolumn{3}{l}{2MASS J17054834-0516462}\\
\noalign{\smallskip}
\hline\noalign{\smallskip}                  
Pleiades MG & $-12,-21,-11$ & $\sim100$ \\
\hline
\noalign{\smallskip}
\multicolumn{3}{l}{2MASS J06023045+3910592}\\
\noalign{\smallskip}
\hline\noalign{\smallskip}                  
AB Doradus MG & $-8,-27,-14$ & $\sim50$ \\
\hline
\noalign{\smallskip}
\multicolumn{3}{l}{2MASS J17312974+2721233}\\
\noalign{\smallskip}
\hline
\end{tabular}
\begin{list}{}{}
\item[$^{\mathrm{a}}$] $UVW$ and age from \citet[][]{Zuckerman04}.
\item[$^{\mathrm{b}}$] Membership already suggested by \citet{Jameson08}, see also Sect.~\ref{sec:notes}
\end{list}
\end{table}

\subsection{Kinematic age of the sample}
\label{sec:age}

\subsubsection{General considerations}
The computation of kinematic ages is based on the dispersion of true
space motions and requires a complete set of 3D velocity information.
The translation of the total velocity dispersion
($\sigma_\mathrm{tot}$) to tangential velocities by means of
$\sigma_\mathrm{tot}=(3/2)^{1/2}\sigma_\mathrm{tan}$ assumes that the
dispersions in $v_\mathrm{tan}$ are spread equally between $UVW$. Even
for a large sample of objects, this assumption might introduce a bias
in the age determination from tangential velocities alone, and
radial velocity measurements are needed to complete the full set of 
velocity vectors.

The velocity dispersion of a statistically significant sample of stars
allows an estimate of the age of the sample, given the dynamical
evolution of the Galaxy (see, e.g., \citealt{Wielen77} or
\citealt{Fuchs01}). Examples of age estimations for samples
including L dwarfs are given in \citet{Schmidt07}, \citet{Zapatero07},
and \citet{Faherty09}. While the results of \citet{Schmidt07} and
\citet{Faherty09} are based on large samples of tangential velocities
only, \citet{Zapatero07} present radial velocities and thus true space
motions of 21 L and T dwarfs. We revisit the population of early- to
mid-L dwarfs and estimate its age based on the total velocity
dispersion of the space motions in our larger sample.

To compute the age of a sample of stars based on the work of
\citet{Wielen77}, one has to calculate the velocity dispersion in $U$,
$V$, and $W$ according to Eqs.~1--3 in \citet{Wielen77}, i.e., one must
apply a weight of $|W|$ to all data points when calculating the
dispersions in each dimension. The weights account for the fact that
we are limited to objects within a small volume close to the midplane of
the Galactic disk, and enable us to infer ages from the kinematical 
heating of the whole disk.

We briefly summarise the results from three papers dealing with
the kinematic age of low-mass objects. \citet{Schmidt07} infer an age
estimate of 3.1~Gyr for their sample of M7--L8 L dwarfs, which agrees
with the value reported by \citet{Faherty09} and appears consistent 
with results for late-M samples \citep{Reid02a, RB09}. On
the other hand, \citet{Zapatero07} report an age of only $\sim1$~Gyr
for their sample of L and T dwarfs. We note that an
important difference with respect to our work is that \citet{Schmidt07}, \citet{Zapatero07}, and
\citet{Faherty09} calculate the total velocity dispersion as the 
dispersion of $v_\mathrm{tot}$ instead of $\sigma_\mathrm{tot} =
\left(\sigma_U^2+\sigma_V^2+\sigma_W^2\right)^{1/2}$, which may
underestimate the dispersion and thus the age. Unfortunately, an error
introduced in the formula computing age from velocity spread had the
effect of almost eliminating this offset, so that in two of the papers,
an age of 3\,Gyr is computed.\footnote{This can be illustrated when
  considering, e.g., the total velocity dispersions in Table~5 of
  \citet{Zapatero07}. For G-type stars, the dispersion peaks at
  $\sigma_\mathrm{tot}=35.9$~km\,s$^{-1}$, which should yield an age
  lower than $\sim2.2$~Gyr (see Fig.~1 in \citet{Fuchs01} and tables
  in \citet{Wielen77}). An age of 9\,Gyr is reported in Table~5 of
  \citet{Zapatero07}.} A more detailed discussion of the
computations of kinematic age by \citet{Schmidt07}, \citet{Zapatero07}, and
\citet{Faherty09} is given in \citet{RB09}. 

We reanalyse the space motions of the sample of 21 brown dwarfs of
spectral type L4V--T8V of \citet{Zapatero07} and calculate $|W|$-weighted 
velocity dispersions of $\left(\sigma_U,\sigma_V,\sigma_W\right) =
(35.0,15.1,15.3)$~km\,s$^{-1}$ from the values given in Tables 3 and 4
of their paper. 
The $|W|$-weighted total velocity dispersion is 
$\sigma_\mathrm{tot}=41.1$~km\,s$^{-1}$. This dispersion
is an indicative of an age of $\sim3.0$~Gyr, which is in good agreement
with ages derived for late-M dwarfs. \citet{Reid02a} infer an age of
3~Gyr from the velocity distribution of a sample of 31 M7.5--M9
dwarfs, and \citet{RB09} find the same age for a
volume-limited sample of 63 M7.5--M9 dwarfs in the solar
neighbourhood.

\subsubsection{Kinematic age of our sample}

We now turn to our new sample of L dwarfs; we find that the velocity
dispersion of our 43 targets is
$\left(\sigma_U,\sigma_V,\sigma_W\right) =
(33.8,28.0,16.3)$~km\,s$^{-1}$ and calculate a total velocity
dispersion of $\sigma_\mathrm{tot} = 46.8$~km\,s$^{-1}$. Our
$|W|$-weighted velocity dispersions are
$\left(\sigma_U,\sigma_V,\sigma_W\right) =
(39.8,29.7,17.4)$~km\,s$^{-1}$ and the $|W|$-weighted total velocity
dispersion is $\sigma_\mathrm{tot}=52.7$~km\,s$^{-1}$. From this value
and Eq.~(16) of \citet{Wielen77}, we derive an age of $\sim5.1$~Gyr
for our sample.

Our result of $\sim5.1$~Gyr is much higher than the $\sim3$~Gyr we
derived for the 21 L and T dwarfs in the sample of \citet{Zapatero07}.
This value also seems counterintuitive compared to the age of
$\sim3$~Gyr for late-M dwarfs. If anything, one would expect a sample of L0--L8
dwarfs to be younger than a sample of M7.5--M9 dwarfs: the hydrogen-burning 
minimum mass (HBMM) divides late-type objects into low mass
stars and brown dwarfs. Comparing the effective temperatures of the
lowest mass stars at 1--10~Gyr given in \citet{Chabrier97} with the
temperature scale of \citet{Vrba04} for L and T dwarfs, this boundary
occurs at spectral type L1--L2.  Brown dwarfs below the HBMM cool over
time and move to much later spectral types than early-L.  This causes
a depletion of old brown dwarfs with masses just below the HBMM in a
sample of early- to mid-L dwarfs \citep[see][]{Burgasser04}. Given
this depletion of old brown dwarfs with spectral types of early- to
mid-L, the average age of a sample of L0--L8 dwarfs as presented here
might be younger, not older, than a sample of late-M dwarfs.

Since we can expect this to affect early-L dwarfs close to the HBMM differently 
from the mid to late-L dwarfs, we divided our sample in half and computed the 
velocity dispersion and age for objects for spectral types L0V-L1.5V and L2V-L8V 
separately. We found no statistically significant difference between the dispersions
and ages of these two groups, nor did we notice any velocity outliers related to 
only one of the groups. This is consistent with a similar analysis by \citet{Schmidt07}, 
who analyse the tangential velocity of a sample of M7--L8 dwarfs and find no significant 
differences in the velocity dispersion and number of outliers between subsamples
of M7V--L2V and L3V--L8V.

\subsubsection{Potential explanations of the old age}

\begin{figure*}
  \centering \resizebox{\hsize}{!}{
    \includegraphics[]{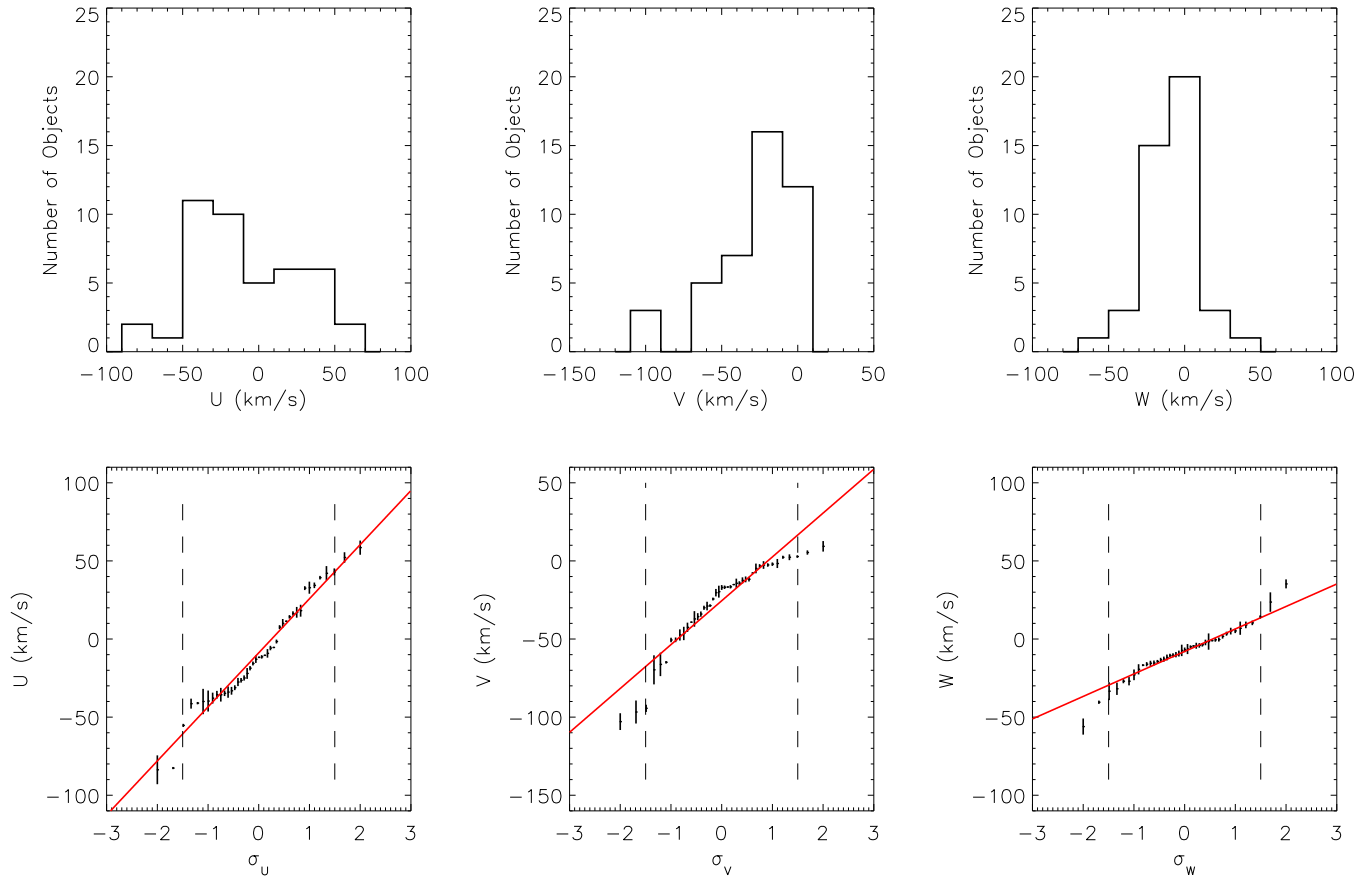}
  }
  \caption{\label{fig:probit}Top row: Histograms of the space motion
    components in $U$, $V$, and $W$ (from left to right). Bottom row:
    Probability plots for in $U$, $V$, and $W$ (from left to right).
    Data are shown with their uncertainties as error bars. The dashed
    line marks the 1.5$\sigma$ cutoff for the fit shown as a solid red
    line. The slope of the linear fit is the $\sigma$ of the closest
    matching Gaussian distribution.}
\end{figure*}

We can speculate about the reason for this result:

(1) The velocity dispersion of our sample might not allow us to
determine an age of our sample because of the non-Gaussianity of
the distribution. While a dispersion can be formally obtained for any
distribution of velocities, the dispersion can only be used as an age
indicator if the distribution is normal. We show the
distributions of $U$, $V$, and $W$ velocities in our sample in the top row
of Fig.~\ref{fig:probit}. In the bottom row of the same figure, we
show probability plots \citep[see, e.g.,][]{Lutz80, Reid02} to help assess
how closely a single Gaussian distribution fits our data. As can be seen
in the plots, a straight line, corresponding to a single Gaussian
distribution, is only marginally consistent with our data. The closest
fit is obtained for the distribution of the $W$ component, while the
$U$ and $V$ components show notable deviations from a Gaussian
distribution. Using our measured velocity dispersion $\sigma_W$ and 
the relation between $\sigma_W$ and $\sigma_\mathrm{tot}$ given in 
\citet{Wielen77}, we compute an age of 3.2 Gyr from $\sigma_W$ alone, 
which is indeed consistent with the age of late-M dwarf samples.

The non-Gaussian distribution of the space motion is probably not
due to the limited size of our sample. \citet{RB09} show
probability plots for their sample of 63 late-M dwarfs, which are very
well reproduced by a single Gaussian distribution. Our sample is only
$\sim30\%$ smaller but exhibits strong deviations from a Gaussian
distribution. The particular choice of the targets for the
sample might instead produce a bias. Since our sample comprises
the brightest objects per spectral bin, the sample should be biased
towards younger objects.  This is, however, not what we found. 
From $\sigma_\mathrm{tot}$, the sample appears much \emph{older}.
On the other hand, the derived age from $\sigma_W$ alone 
is substantially younger than the one inferred from $\sigma_\mathrm{tot}$. In fact, 
the 3~Gyr derived from $\sigma_W$ is similar to the age of the late-M samples, which 
may be expected to be of comparable age. Because the dispersions in $U$ 
and $V$ do not follow a Gaussian distribution while the one in $W$ does, one 
might argue that the age from $W$ alone is a more reliable estimate of the true 
age of our sample. Unfortunately, at this point we have no explanation of
why the distributions in $U$ and $V$ differ from the one in $W$.

(2) The lowest mass stars and brown dwarfs might have a different
initial velocity dispersion at the time of formation from
higher mass stars \citep[$v_0=10$~km\,s$^{-1}$ in][]{Wielen77}.
\citet{Bihain06} find that the velocity dispersion of the brown dwarf
members in the Pleiades is about four times the dispersion of solar-type 
members of the cluster. A higher $v_0$ would naturally lead to a
different velocity dispersion as a function of time without invoking a
different feedback by the lowest mass stars and brown dwarfs on the
Galactic potential. This effect, however, should have identical influences 
on $U$, $V$, and $W$. We see no reason why only $U$ and $V$ should be
affected by such a mechanism.

(3) Individual kinematical outliers can significantly bias the
velocity dispersions of small samples. This effect would be enhanced
by the $|W|$-weighting, as discussed in \citet{Reid02}. However, we
find no clear correlation of $|W|$ with either $v_\mathrm{tot}$ or
$v_\mathrm{tan}$ in our sample, and the removal of five objects
with the highest $v_\mathrm{tot}$ only marginally lowers the age
of the sample. On the other hand, the removal of six objects with
the highest \emph{tangential} velocity ($v_\mathrm{tan} >
70$~km\,s$^{-1}$) lowers the mean age of the sample to 3.0~Gyr,
as computed from $\sigma_\mathrm{tot}$. This effect may be expected but a
removal of those objects can only be justified if the distribution of
tangential velocities identifies them as clear outliers. This is not
evident for our sample. We note that the error in the spectrophotometric 
distances of some targets might be much larger than the $\pm1$\,pc given 
in \citet{Faherty09}, which could contribute to a few moderate $v_\mathrm{tan} $ 
outliers.  In any case, it remains true that even 43 objects must be considered too small a 
sample when robustness to kinematic outliers is important, especially 
when the sample is not volume-limited and prone to selection effects.

In summary, an age of $\sim$3~Gyr for our sample seems likely given 
the non-Gaussian distribution of the $U$ and $V$ velocities and the age 
determined from $\sigma_W$ alone. Furthermore, that the 
removal of the most significant outliers in the distribution of tangential 
velocities reduces the age of the sample to a similar value may indicate 
that a handful of fast moving outliers are contained in the sample. 
These outliers may have been accelerated during a very early phase of their 
evolution, perhaps by ejection from multiple systems.

\subsection{Notes on individual objects}\label{sec:notes}
\begin{itemize}{}{}
\item{\textit{2MASS 0036+18} and \textit{2MASS 0825+21}} are both
  listed in \citet{Bannister07} as belonging to the Hyades Moving Group
  based on the moving cluster method. We are unable to confirm this finding.
  Although both objects belong to the young disk population, their
  space motions are inconsistent with the space vector of the Hyades
  in $V$ and $W$ (2MASS 0036+18) and $U$ and $W$ (2MASS 0825+21),
  respectively.

\item{\textit{2MASS 0255-47}}: We report a radial velocity for this
  object of $v_\mathrm{rad}=25\pm4.1$~km\,s$^{-1}$, which is
  inconsistent with $v_\mathrm{rad}=17.5\pm2.8$~km\,s$^{-1}$ reported
  in \citet{Zapatero07}, and $v_\mathrm{rad}=13.0\pm3.0$~km\,s$^{-1}$
  in \citet{B00}. Given the late spectral type, ranging from L6 to L8,
  and the high rotational velocity of the system
  ($v\sin{i}\approx67$km\,s$^{-1}$, RB08) the uncertainties in all the
  reported radial velocity values -- including ours -- might be
  underestimated. Nevertheless, we note a slight trend with time in the 
  radial velocity that should be verified with additional observations.

\item{\textit{2MASS 1045-01} and \textit{2MASS 1047-18}}:
  \citet{Jameson08} list both objects as probable Hyades members,
  based on new proper motion measurements. We can confirm their
  membership based on the full space motions of both objects.

\item{\textit{2MASS 1305-25}}: This object, also known as Kelu 1, is a
  tight binary \citep[$\rho\sim0.3$\arcsec,][]{Liu05} consisting of a
  L1.5-L3 and a L3-L4.5 component. Its orbit is highly eccentric and
  the orbital period is estimated to be $\sim38$~yrs \citep{Stumpf08}. 
  Using the orbital elements of \citet{Stumpf08}, we estimate that our 
  radial velocity measurement might be affected by the orbital motion
  of the binary at a level of a few km\,s$^{-1}$ depending on the
  fractional light contribution from each component in our optical
  spectra. Because of the high rotational velocity of the components in
  the system ($v\sin{i}\approx76$~km\,s$^{-1}$, RB08), we are unable to
  identify the individual components in the spectrum. We note that
  \citet{B00} report a radial velocity of
  $v_\mathrm{rad}\sim17$~km\,s$^{-1}$ measured in June 1997. This
  value differs significantly from the
  $v_\mathrm{rad}=5\pm2.2$~km\,s$^{-1}$ that we measure in our
  spectrum from May 2006, although the difference seems too large to
  be explained by the orbital motion only.

\item{\textit{2MASS 1441-09}}: \citet{Bouy03} found this object to
  be a close binary ($\rho\sim0.4$\arcsec) composed of two L1 dwarfs.
  \citet{Bouy08} present follow-up astrometric observations and
  estimate the orbital period to be 120-230~yr. This period is too long
  and the brightness difference in the two components too small to
  bias our radial velocity measurement in any significant way.
  \citet{Seifahrt05} demonstrate that this L dwarf binary forms a
  common proper motion pair with the M4.5V star G124-62. The radial
  velocity of G124-62 reported by the same authors is
  $v_\mathrm{rad}=-29.3$~km\,s$^{-1}$, which is in close agreement
  with our radial velocity for the L dwarf binary. \citet{Seifahrt05}
  argue that the system is a probable member of the Hyades
  supercluster, which we confirm with improved proper motion for the
  system.

\item{\textit{2MASS 1854+84}}: This object has no published proper
  motion and distance. Hence, we have to exclude it from our analysis
  of space motions, reducing the original sample size to 43 objects.
  We obtain a radial velocity of
  $v_\mathrm{rad}=-3.5\pm1.0$~km\,s$^{-1}$ for this target.

\item{\textit{2MASS 2200-30}}: \citet{Burgasser06} resolve this target
  into a close binary ($\rho\sim1$\arcsec) with spectral types of
  M9V+L1V and estimate the orbital period to be 750--1000~yrs.
  Hence, our radial velocity measurement should be unaffected by the
  binarity, given the long period and similar brightness of both
  components.

\end{itemize}

\section{Summary}\label{sec:summary}
We have measured the radial velocities of 44 L0--L8 dwarfs in the solar
neighbourhood from high resolution optical VLT/UVES and Keck/HIRES
spectra. We combined these measurements with distance and proper
motion data to compute space motions for 43 of our targets. For one
target (2M1854+84), no proper motion and distance was available in the
literature and we had to exclude the target from the sample. We
formally derived an age of 5.1~Gyr from the $|W|$-weighted total
velocity dispersion of the sample and the age-velocity relation of
\citet{Wielen77}. However, the space motion components $U$ and $V$ of
our sample are not normally distributed, compromising the age
computation from $\sigma_\mathrm{tot}$. We found that the
distribution in $W$ is, however, quite consistent with a Gaussian shape. 
From $W$ alone, we compute an age of $\sim3$~Gyr.

We note that the formally derived age of 5.1~Gyr from $\sigma_\mathrm{tot}$
for the whole sample is in disagreement with the mean age of
the lowest mass stars in the solar neighbourhood
\citep[$\sim3$~Gyr,][]{Reid02a,RB09} and the expectation that
early- to mid-L dwarfs should be younger than the lowest mass
stars of spectral type M7.5--M9. We speculate that either the initial
velocity dispersion of brown dwarfs is notably different from that
of solar-type and lower-mass stars, or that our particular sample is
biased towards older and/or kinematically peculiar objects. The latter
seems the most likely explanation since a few high-velocity objects
could bias our statistics.  Such a bias would be enhanced by the
$|W|$-weighting of the dispersions when calculating
$\sigma_\mathrm{tot}$ and thus the mean age of our sample.

The most likely age of nearby early- to mid-L type dwarfs is of the
order of 3\,Gyr, and we have found no evidence of an age lower than this. 
However, a statistically clearly defined, and perhaps
larger sample of L and T dwarfs with accurate radial velocities would
be necessary to robustly determine the mean age of brown dwarfs from a
kinematic analysis.

\begin{acknowledgements}

  This work is based on observations obtained from the European
  Southern Observatory, PIDs 077.C-0449 and 078.C-0025, and on
  observations obtained from the W.M. Keck Observatory, which is
  operated as a scientific partnership among the California Institute
  of Technology, the University of California and the National
  Aeronautics and Space Administration. Based on observations made
  with the European Southern Observatory telescopes obtained from the
  ESO/ST-ECF Science Archive Facility. This research has benefitted 
  from the M, L, and T dwarf compendium housed at DwarfArchives.org 
  and maintained by Chris Gelino, Davy Kirkpatrick, and Adam Burgasser. 
  We thank the referee, Kelle Cruz, for her constructive report.
  AS and AR acknowledge financial support 
  from the Deutsche Forschungsgemeinschaft under DFG
  RE 1664/4-1. AS further acknowledges financial support from NSF grant 
  AST07-08074. GB thanks the NSF for grant support through AST06-06748.
\end{acknowledgements}

\bibliographystyle{aa}

\begin{landscape}
\begin{table*}

\caption{Radial velocities and space motions of our targets. Data on distance and proper motions are obtained from the literature.}
\label{maintable}
\hspace*{-5cm}
\begin{tabular}{llr@{ $\pm$ }lr@{ $\pm$ }rr@{ $\pm$ }rr@{ $\pm$ }lr@{ $\pm$ }lr@{ $\pm$ }lr@{ $\pm$ }lcl} 
\hline\hline             
Object & SpT & \multicolumn{2}{c}{Distance}  & \multicolumn{2}{c}{$\mu_{\alpha}\cos{\delta}$} &  \multicolumn{2}{c}{$\mu_{\delta}$} & \multicolumn{2}{c}{RV} & \multicolumn{2}{c}{$U$} & \multicolumn{2}{c}{$V$} & \multicolumn{2}{c}{$W$} & Ref. & Comments\\
       &     & \multicolumn{2}{c}{(pc)}  & \multicolumn{2}{c}{(mas/yr)} &  \multicolumn{2}{c}{(mas/yr)} & \multicolumn{2}{c}{(km\,s$^{-1}$)} & \multicolumn{2}{c}{(km\,s$^{-1}$)} & \multicolumn{2}{c}{(km\,s$^{-1}$)} & \multicolumn{2}{c}{(km\,s$^{-1}$)} &   &  \\
\hline
2MASS J03140344+1603056 & L0.0 & 14.0 & 1.0 &  -241 & 18 &   -76 & 19 &  -8.0 & 1.1 &   16.4 & 1.4 &    5.4 & 1.4 &   -7.0 & 1.5 & 1 &     UMa moving group \\
2MASS J12212770+0257198 & L0.0 & 19.0 & 1.0 &  -115 & 30 &   -18 & 27 &  -9.0 & 1.4 &   -9.2 & 2.8 &   -2.6 & 2.5 &  -10.0 & 1.7 & 1 &           Young disk \\
2MASS J17312974+2721233 & L0.0 & 12.0 & 1.0 &   -82 & 15 &  -240 & 17 & -30.5 & 1.1 &   -5.8 & 1.4 &  -30.0 & 1.3 &  -14.3 & 1.0 & 1 & AB Dor moving group  \\
2MASS J22000201-3038327 & L0.0 & 35.0 & 2.0 &   210 & 48 &   -64 & 21 & -25.3 & 1.0 &  -39.7 & 6.7 &  -19.7 & 3.8 &   -1.6 & 5.2 & 1 & Hyades moving group  \\
2MASS J07464256+2000321 & L0.5 & 12.2 & 0.04 &  -374 &  1 &   -57 &  0 &  53.0 & 1.1 &  -55.3 & 1.0 &  -15.1 & 0.4 &   -1.0 & 0.4 & 1 &                      \\
2MASS J14122449+1633115 & L0.5 & 25.0 & 2.0 &    29 & 16 &   -80 & 30 &   5.0 & 1.1 &   10.0 & 2.8 &   -4.8 & 3.1 &    1.7 & 1.5 & 1 &           Young disk \\
2MASS J14413716-0945590 & L0.5 & 27.5 & 2.7 &  -198 &  3 &   -16 &  4 & -28.3 & 1.1 &  -34.7 & 1.8 &  -12.5 & 2.0 &  -10.4 & 1.3 & 1 & Hyades moving group  \\
2MASS J23515044-2537367 & L0.5 & 18.0 & 2.0 &   387 & 21 &   163 &  9 &  -3.0 & 1.1 &  -35.7 & 4.4 &   -2.1 & 1.1 &   -4.1 & 1.4 & 1 &           Young disk \\
2MASS J02355993-2331205 & L1.0 & 21.3 & 0.5 &    84 &  1 &    13 &  0 &  15.3 & 1.1 &  -11.7 & 0.4 &   -7.7 & 0.3 &  -10.5 & 1.1 & 2 &           Young disk \\
2MASS J06023045+3910592 & L1.0 & 10.6 & 0.3 &   146 & 10 &  -501 & 10 &   7.6 & 1.1 &  -11.4 & 1.1 &  -24.4 & 0.9 &   -4.4 & 0.6 & 1 & Pleiades moving group \\
2MASS J10224821+5825453 & L1.0 & 20.0 & 1.0 &  -360 & 99 &  -780 & 90 &  19.7 & 1.1 &  -40.0 & 8.1 &  -69.7 & 9.3 &   23.7 & 6.2 & 1 &                      \\
2MASS J10452400-0149576 & L1.0 & 17.0 & 1.0 &  -495 & 18 &    12 & 12 &   7.0 & 1.1 &  -35.7 & 2.5 &  -14.2 & 1.3 &  -12.9 & 1.6 & 1 & Hyades moving group \\
2MASS J10484281+0111580 & L1.0 & 15.0 & 1.0 &  -442 & 13 &  -209 & 12 &  24.0 & 1.1 &  -24.9 & 1.6 &  -34.0 & 1.7 &   -2.2 & 1.8 & 1 &                      \\
2MASS J13004255+1912354 & L1.0 & 14.0 & 1.0 &  -820 & 18 & -1244 & 19 & -17.8 & 1.0 &   -1.6 & 1.3 &  -96.8 & 7.4 &  -27.0 & 1.3 & 1 &                      \\
2MASS J13595510-4034582 & L1.0 & 21.0 & 1.0 &    38 & 11 &  -485 & 15 &  49.8 & 1.0 &   39.3 & 1.1 &  -50.5 & 1.5 &  -27.0 & 2.6 & 1 &                      \\
2MASS J14392836+1929149 & L1.0 & 14.4 & 0.1 & -1229 &  1 &   407 &  2 & -27.2 & 1.1 &  -82.6 & 0.7 &  -39.2 & 0.3 &   14.1 & 1.1 & 1 &                      \\
2MASS J15551573-0956055 & L1.0 & 13.0 & 1.0 &   950 & 15 &  -767 & 15 &  14.5 & 1.1 &   52.2 & 3.4 &    2.3 & 1.0 &  -56.1 & 5.2 & 1 &                      \\
2MASS J11455714+2317297 & L1.5 & 44.0 & 3.0 &   155 & 16 &   -56 &  6 &   3.3 & 1.0 &   32.9 & 3.9 &    2.4 & 1.9 &   10.3 & 1.4 & 1 &                      \\
2MASS J13340623+1940351 & L1.5 & 46.0 & 3.0 &   -58 & 12 &    98 & 16 &   2.0 & 4.1 &  -22.0 & 3.5 &    9.3 & 3.4 &    6.9 & 4.2 & 1 &                      \\
2MASS J16452211-1319516 & L1.5 & 12.0 & 1.0 &  -364 & 18 &  -804 & 16 &  26.4 & 1.0 &   32.6 & 1.2 &  -46.4 & 4.3 &   -0.3 & 1.3 & 1 &                      \\
2MASS J18071593+5015316 & L1.5 & 14.0 & 1.0 &    35 & 19 &  -126 & 14 &  -2.0 & 3.2 &    7.6 & 1.3 &   -1.7 & 2.9 &   -4.3 & 1.9 & 1 &           Young disk \\
2MASS J20575409-0252302 & L1.5 & 16.0 & 1.0 &    10 & 20 &   -80 & 20 & -25.0 & 3.2 &  -12.9 & 2.3 &  -20.3 & 2.4 &    9.0 & 2.1 & 1 &           Young disk \\
2MASS J08283419-1309198 & L2.0 & 11.6 & 1.4 &  -592 &  6 &    14 &  7 &  26.2 & 1.0 &  -33.1 & 2.5 &  -16.6 & 1.1 &  -19.4 & 3.3 & 3 &           Young disk \\
2MASS J09211410-2104446 & L2.0 & 12.0 & 1.0 &   244 & 16 &  -908 & 17 &  80.0 & 1.1 &   18.1 & 3.8 &  -94.4 & 2.3 &    5.1 & 2.2 & 1 &                      \\
2MASS J11553952-3727350 & L2.0 & 13.0 & 1.0 &    50 & 12 &  -767 & 15 &  45.0 & 1.1 &   34.4 & 1.8 &  -50.6 & 1.5 &  -22.9 & 3.4 & 1 &                      \\
2MASS J13054019-2541059 & L2.0 & 18.7 & 0.7 &  -285 &  1 &    11 &  1 &   5.0 & 2.2 &  -18.7 & 1.4 &  -17.0 & 1.6 &    5.2 & 1.4 & 1 &           Young disk \\
2MASS J05233822-1403022 & L2.5 & 13.0 & 1.0 &    90 & 17 &   166 & 17 &  11.3 & 1.0 &  -15.7 & 1.2 &   -2.6 & 1.1 &    3.4 & 1.3 & 1 &           Young disk \\
2MASS J10292165+1626526 & L2.5 & 23.0 & 2.0 &   359 & 15 &  -348 & 17 & -28.2 & 1.4 &   58.5 & 4.5 &  -14.3 & 2.9 &  -11.8 & 1.9 & 1 &                      \\
2MASS J10473109-1815574 & L2.5 & 22.0 & 2.0 &  -347 & 17 &    54 & 14 &   6.0 & 1.1 &  -34.0 & 3.6 &  -11.7 & 1.5 &   -9.3 & 1.9 & 1 & Hyades moving group \\
2MASS J09130320+1841501 & L3.0 & 46.0 & 4.0 &    32 & 16 &  -187 & 17 &  -1.0 & 2.2 &   17.1 & 3.5 &  -37.1 & 5.0 &   -6.4 & 3.2 & 1 &                      \\
2MASS J12035812+0015500 & L3.0 & 19.0 & 2.0 & -1209 & 18 &  -261 & 15 &  -1.5 & 2.2 &  -83.7 & 9.2 &  -66.3 & 7.5 &  -31.9 & 3.9 & 1 &                      \\
2MASS J15065441+1321060 & L3.0 & 14.0 & 1.0 & -1087 & 13 &    14 & 11 &  -0.9 & 1.2 &  -41.4 & 3.2 &  -47.4 & 3.6 &   35.3 & 2.9 & 1 &                      \\
2MASS J16154416+3559005 & L3.0 & 24.0 & 2.0 &   -17 & 12 &  -512 & 15 & -21.0 & 1.2 &   42.1 & 4.6 &  -42.5 & 3.0 &  -16.1 & 1.3 & 1 &                      \\
2MASS J21041491-1037369 & L3.0 & 17.0 & 2.0 &   614 & 16 &  -281 & 15 & -20.5 & 2.2 &  -37.9 & 3.5 &  -28.9 & 2.7 &  -33.4 & 5.7 & 1 &                      \\
2MASS J00361617+1821104 & L3.5 &  8.8 & 0.1 &   899 &  1 &   120 &  2 &  21.0 & 1.8 &  -41.0 & 0.7 &   -3.2 & 1.2 &  -13.3 & 1.3 & 1 &           Young disk \\
2MASS J07003664+3157266 & L3.5 & 12.2 & 0.3 &   130 &  1 &  -546 &  1 & -43.5 & 2.2 &   43.0 & 2.2 &  -28.6 & 0.8 &  -16.8 & 0.6 & 1 &                      \\
2MASS J17054834-0516462 & L4.0 & 11.0 & 1.0 &   129 & 14 &  -103 & 15 &  12.2 & 1.1 &   14.2 & 1.1 &    2.8 & 0.8 &   -3.7 & 1.1 & 1 &     UMa moving group \\
2MASS J22244381-0158521 & L4.5 & 11.4 & 0.1 &   457 &  2 &  -871 &  1 & -39.0 & 1.1 &  -10.4 & 0.4 &  -64.9 & 0.8 &   -5.0 & 0.9 & 1 &                      \\
2MASS J00043484-4044058 & L5.0 & 13.0 & 0.7 &   643 &  2 & -1494 &  1 &  30.0 & 1.4 &   11.4 & 0.5 & -102.9 & 5.4 &  -15.3 & 1.6 & 4 &                      \\
2MASS J08251968+2115521 & L5.0 & 10.7 & 0.1 &  -510 &  2 &  -288 &  2 &  20.0 & 3.2 &  -27.5 & 2.6 &  -16.9 & 1.1 &  -15.5 & 1.6 & 1 &           Young disk \\
2MASS J08354256-0819237 & L5.0 &  7.0 & 1.0 &  -530 & 30 &   300 & 30 &  26.5 & 2.2 &  -31.2 & 1.6 &  -11.8 & 1.9 &   -0.5 & 1.2 & 1 & Hyades moving group \\
2MASS J15074769-1627386 & L5.0 &  7.3 & 0.03 &  -161 &  2 &  -888 &  0 & -40.4 & 1.8 &  -26.4 & 1.5 &  -16.7 & 0.4 &  -40.5 & 1.1 & 1 &                      \\
2MASS J02550357-4700509 & L8.0 &  5.0 & 0.1 &   999 &  2 &  -565 &  3 &  25.0 & 4.1 &   -5.4 & 0.4 &  -35.7 & 2.2 &   -7.3 & 3.6 & 5 &                     \\                                  
\hline
\end{tabular}

\begin{list}{}{}
\item[\textsc{References. -- }]Distances and proper motion: (1) \citet[][and references therein]{Faherty09}, 
(2) HIPPARCOS catalogue\\ \citep{Perryman97}, (3) \citet{Lodieu05}, (4) \citet{Henry06}, (5) \citet{Costa06}
\end{list}

\end{table*}
\end{landscape}

\end{document}